\documentclass{PoS}

\usepackage{array}

\title{Symmetry energy and density}

\ShortTitle{Symmetry energy and density}

\author{\speaker{Wolfgang TRAUTMANN}\\%
        GSI Helmholtzzentrum Gmbh, D-64291 Darmstadt, Germany\\
        E-mail: \email{w.trautmann@gsi.de}}

\author{Mircea Dan COZMA\\
        IFIN-HH, Reactorului 30, 077125 M\v{a}gurele-Bucharest, Romania\\
        E-mail: \email{dan.cozma@theory.nipne.ro}}

\author{Paolo RUSSOTTO\\
        INFN-Sezione di Catania, I-95123 Catania, Italy\\
        E-mail: \email{russotto@lns.infn.it}}

\abstract{The nuclear equation-of-state is a topic of highest current interest in nuclear structure and reactions as well as in astrophysics. 
In particular, the equation-of-state of asymmetric matter and the symmetry energy representing the difference between the energy densities 
of neutron matter and of symmetric nuclear matter are not sufficiently well constrained at present. 
The density dependence of the symmetry energy is conventionally expressed in the form of the slope parameter $L$ describing the derivative 
with respect to density of the symmetry energy at saturation. Results deduced from nuclear structure and heavy-ion reaction data are
distributed around a mean value $L=60$~MeV. 

Recent studies have more thoroughly investigated the density range that a particular observable is predominantly sensitive to. 
Two thirds of the saturation density is a value typical for the information contained in nuclear-structure data. Higher values 
exceeding saturation have been shown to be probed with meson production and collective flows at incident energies in the range of 
up to about 1 GeV/nucleon.

From the measurement of the elliptic-flow ratio of neutrons with respect to light charged particles in recent experiments at 
the GSI laboratory, a new more stringent constraint for the symmetry energy at suprasaturation
density has been deduced. It confirms, with a considerably smaller uncertainty, the moderately soft to linear density dependence 
of the symmetry energy previously deduced from the FOPI-LAND data. Future opportunities offered by FAIR will be discussed.
}

\FullConference{54th International Winter Meeting on Nuclear Physics\\
		25-29 January 2016 \\
		Bormio, Italy}

\begin{document}


\section{Introduction}

In a recent paper, Li and Han have documented that the many results obtained for the nuclear symmetry energy from terrestrial nuclear experiments and 
astrophysical observations are amazingly compatible, even though individual results scatter within certain margins and are partly  
affected with considerable errors~\cite{lihan2013}. The average values deduced by the authors for the symmetry energy at saturation density and for the
slope parameter describing its density dependence are $E_{\rm sym}(\rho_0) = 31.6$~MeV and $L = 58.9$~MeV, respectively. 
The parameter $L$, defined as
\begin{equation}
L = 3\rho_0 \frac{\partial E_{\rm sym}}{\partial \rho}|_{\rho=\rho_0}, 
\label{eq:L}
\end{equation}
is proportional to the slope of the symmetry energy at saturation density $\rho_0$ (see, e.g., Ref.~\cite{lipr08}). 
The authors quote also error margins representative for the variation of the individual results as 
$\Delta E_{\rm sym}(\rho_0) = 2.7$~MeV and $\Delta L = 16$~MeV and conclude that $L$ has a value about twice as large
as $E_{\rm sym}(\rho_0)$. A very similar conclusion can be drawn from the compilation of Lattimer and Steiner~\cite{lattimer14} adapted from
the earlier work of Lattimer and Lim~\cite{lattlim13}. 

The nuclear symmetry energy appears in many aspects of nuclear structure and reactions and determines very basic properties of neutron stars as, e.g., 
their radii~\cite{lipr08,epja2014}. 
This implies that many different sources of information exist, in the laboratory and in the cosmos, that provide information on the equation of 
state (EoS) of asymmetric nuclear matter. Comparisons of the mentioned kind are possible because the forces identified as best describing a particular observation
can be used in many-body calculations for generating results for the two quantities characterizing the asymmetric-matter EoS at saturation density. 
It does not require that the measurement or observation has actually tested the EoS at this density. The result may represent an extrapolation.  

The predictions of microscopic models for the nuclear symmetry energy, obtained with realistic or phenomenological forces, appear to coincide at densities
close to $\rho = 0.1$~fm$^{-3}$, i.e. at approximately two thirds of the saturation density~\cite{fuchs06}. It reflects the fact that the average density of 
atomic nuclei is below saturation and that the presence of the surface influences the nuclear properties that are chosen as constraints. The awareness that each 
observable carrying information on the nuclear EoS is connected to its proper range of density has to complement the interpretation of existing results. 

\begin{figure}[htb]           
\centering
\includegraphics[width=8.0cm]{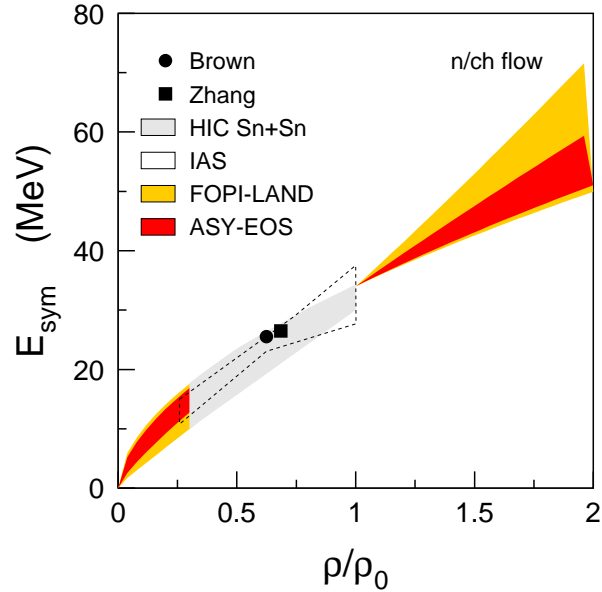}
\caption{(Color online) Constraints deduced for the density dependence of the symmetry energy from the ASY-EOS data~\protect\cite{russotto16}
in comparison with the FOPI-LAND result of Ref.~\protect\cite{russotto11} as a function of the reduced density $\rho/\rho_0$. 
The low-density results of Refs.~\protect\cite{brown13,zhang13,tsang09,dani14} as reported in Ref.~\protect\cite{horowitz13} are 
given by the symbols, the grey area (HIC), and the dashed contour (IAS). For clarity, the FOPI-LAND and ASY-EOS results are
not displayed in the interval $0.3 < \rho/\rho_0 < 1.0$
(from Ref.~\protect\cite{russotto16}; Copyright (2016) by the American Physical Society).
}
\label{fig:final}       
\end{figure}

Rather precise values for the symmetry energy and for the density to which it applies have recently
been presented by Brown~\cite{brown13} and Zhang and Chen~\cite{zhang13}. They are shown in Fig.~\ref{fig:final} together with the low-density behavior of 
the symmetry energy obtained from heavy-ion collisions~\cite{tsang09} and from the analysis of isobaric analog states~\cite{dani14} as reported in Ref.~\cite{horowitz13}. 
In the study of Brown, a set of selected Skyrme forces is used whose parameters are fitted to properties of doubly magic nuclei. By using particular 
values for the neutron skin of $^{208}$Pb nuclei within a given range as additional constraints, new sets of of these forces with slightly adjusted 
parameters are obtained. It is found that all predictions coincide at a density $\rho = 0.1$~fm$^{-3}$, independent of the choice made for the 
thickness of the neutron skin, but that the slopes at this density depend on this choice. 
Only a precise knowledge of the neutron skin of $^{208}$Pb will permit the extrapolation 
to the saturation point. It underlines the importance of this quantity determined by the balance of pressures felt by neutrons in the 
neutron-enriched interior of the $^{208}$Pb nucleus at approximately saturation density and in the low-density 
neutron-rich surface~\cite{rocamaza11,prex12,kumar12}.

Sensitivities to higher densities can be expected from observables related to the early phases of heavy-ion collisions at sufficiently high energies. 
Calculations predict that densities up to three times the saturation value are reached for short times ($\approx 20$~fm/$c$) in the central zone of 
heavy-ion collisions with energies of up to $\approx 1$~GeV/nucleon~\cite{xu13}. 
Twice the saturation value may already be reached at the moderate energy of 400 MeV/nucleon at which the 
reaction dynamics is still largely determined by the nuclear mean field. The resulting pressure produces a collective
outward motion of the compressed material whose strength will
be influenced by the symmetry energy in asymmetric systems~\cite{dani02}. A measurement differentiating between the collective flows of 
neutrons and protons can thus be expected to provide information on the high-density symmetry energy~\cite{li02}.

The ratio of elliptic flow strengths observed for neutrons and light charged particles has been proposed as an observable 
sensitive to the EoS of asymmetric matter~\cite{russotto11}. From a reanalysis of the earlier 
FOPI-LAND data~\cite{leif93,lamb94} and the comparison with calculations performed with the UrQMD transport 
model~\cite{qli05,qli06,Li:2006ez} a moderately soft to linear symmetry term, characterized by a coefficient $\gamma = 0.9 \pm 0.4$ 
for the power-law parametrization of the density dependence of the potential part of the symmetry energy was 
obtained~\cite{russotto11,cozma11,traut12,traut14,russotto_epja14}. 
It is represented by the yellow band in Fig.~\ref{fig:final}. Motivated by this finding, an attempt has been made to improve the 
accuracy with a new experiment that was conducted at the GSI laboratory in 2011 (ASY-EOS experiment S394). 
The new result, reported in Ref.~\cite{russotto16}, is represented by the red band in the figure. Both results are displayed over 
the range of densities up to twice saturation. It will be shown in this talk that the sensitivity of this particular observable 
reaches even beyond that point.

\section{The ASY-EOS experiment}
\label{sec:exp}

\begin{figure}[htb]            
\centering
\includegraphics[width=10.5cm]{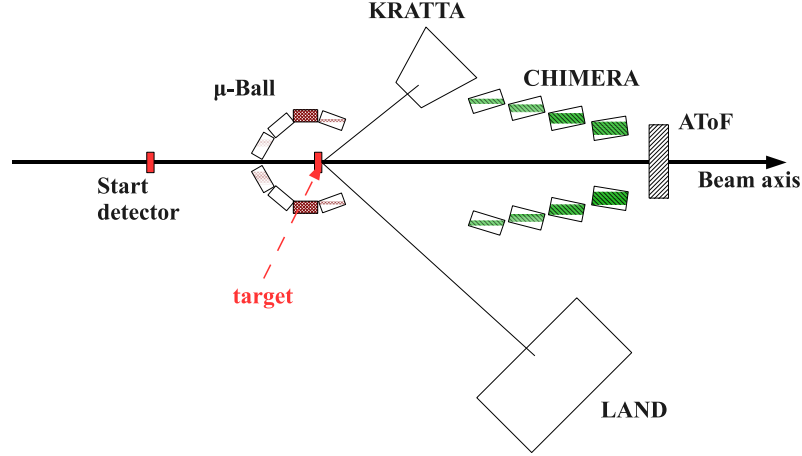}
\caption{(Color online) Schematic view of the experimental setup used in the ASY-EOS
experiment S394 at the GSI laboratory showing the six main detector systems and their 
positions relative to the beam direction. The dimensions of the symbols 
and the distances are not to scale
(from Ref.~\protect\cite{russotto15}).
}
\label{fig:setup}       
\end{figure}

The experimental setup of the ASY-EOS experiment at the GSI laboratory followed the scheme developed for FOPI-LAND by using the Large Area Neutron Detector
(LAND~\cite{LAND}) as the main instrument for neutron and charged particle detection~(Fig.~\ref{fig:setup}). Several detection systems with azimuthal symmetry 
with respect to the beam axis served in determining the orientation of the reaction plane. Upstream from the target, 
a thin plastic scintillator foil viewed by two
photomultipliers was used to record the projectile arrival times and to serve as a start 
detector for the time-of-flight measurement.
LAND was positioned at a laboratory angle close to 
45$^{\circ}$ with respect to the beam direction. A veto wall of thin plastic 
scintillators in front of LAND was used for discriminating between neutrons and charged particles. 
This configuration permitted the measurement of directed and elliptic flows of neutrons and 
charged particles near mid-rapidity within the same angular acceptance.
Opposite of LAND, covering a comparable range of polar angles, the Krak\'{o}w Triple Telescope 
Array (KRATTA~\cite{Luk11}) was installed to permit flow measurements of identified 
charged particles under the same experimental conditions. 

For the event characterization and for measuring the orientation of the reaction plane, three
detection systems had been installed. The ALADIN Time-of-Flight (AToF) Wall~\cite{schuettauf96} was used 
to detect charged particles and fragments in forward direction at polar angles up to
$\theta_{\rm lab} \le 7^{\circ}$. Its capability of identifying large fragments and of characterizing
events with a measurement of $Z_{\rm bound}$~\cite{schuettauf96} permitted the sorting of events 
according to impact parameter. Four double rings of the CHIMERA multidetector~\cite{Pag04,DeFilippo14} 
carrying together 352 CsI(Tl) scintillators in forward direction and four rings with 50 thin CsI(Tl) 
elements of the Washington University Microball array~\cite{muball} surrounding the target
provided sufficient coverage and granularity for determining the orientation 
of the reaction plane from the measured azimuthal particle distributions. A detailed description of the experiment is available in Ref.~\cite{russotto16}.

\section{Experimental results}
\label{sec:urqmd}

As in the FOPI-LAND experiment, the reaction $^{197}$Au+$^{197}$Au at 400 MeV/nucleon was studied. 
The elliptic flows of neutrons and light charged particles were determined from the azimuthal distributions of these particles with 
respect to the reaction plane reconstructed from the distributions of particles and fragments recorded with the three arrays AToF, CHIMERA, and Microball. 
Methods developed and described in Refs.~\cite{Ollxx,andronic06} for correcting the finite dispersion of the reaction plane were applied. 
The coefficients $v_1$ and $v_2$ representing the strengths of directed and elliptic flows, respectively, were deduced from fits with the Fourier expansion 
\begin{equation}
f(\Delta\phi) \propto 1 + 2 v_1 {\rm cos}(\Delta\phi) + 2 v_2 {\rm cos}(2 \Delta\phi).
\label{eq:fourier}
\end{equation}
Here $\Delta\phi$ represents the azimuthal angle of the momentum vector of an 
emitted particle with respect to the angle representing the azimuthal orientation of the reaction plane. 
Constraints for the symmetry energy were determined by comparing the ratios 
of the elliptic flows of neutrons and charged particles (ch), $v_2^{n}/v_2^{ch}$, with the corresponding UrQMD 
predictions for soft and stiff assumptions.

The UrQMD model was originally developed to study particle production at high 
energy~\cite{bass98}. By introducing a nuclear mean field 
with momentum dependent forces, it has been adapted to the study of intermediate energy heavy-ion collisions~\cite{qli09}.
The updated Pauli-blocking scheme, introduced to provide a more precise description of experimental observables 
at lower energies, is described in Ref.~\cite{qli11}. The chosen isoscalar EoS is soft and
different options for the dependence on isospin asymmetry were implemented. Two of them are used here, 
expressed as a power-law dependence of the potential part of the symmetry energy on the
nuclear density $\rho$ according to
\begin{equation}
E_{\rm sym} = E_{\rm sym}^{\rm pot} + E_{\rm sym}^{\rm kin} 
= 22~{\rm MeV} (\rho /\rho_0)^{\gamma} + 12~{\rm MeV} (\rho /\rho_0)^{2/3} 
\label{eq:pot_term}
\end{equation}
with $\gamma =0.5$ and $\gamma =1.5$ corresponding to a soft and a stiff density 
dependence, respectively.  

\begin{figure}[t]           
\centering
\includegraphics[width=8.0cm]{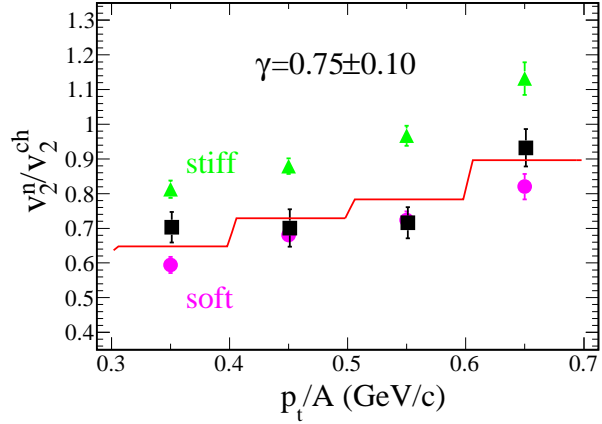}
\caption{(Color online) Elliptic flow ratio of neutrons over all charged particles for central ($b<$ 7.5 fm) 
collisions of $^{197}$Au+$^{197}$Au at 400 MeV/nucleon
as a function of the transverse momentum/nucleon $p_{t}/A$. 
The black squares represent the experimental data, the green triangles and purple circles represent the UrQMD predictions
for stiff ($\gamma =1.5$) and soft ($\gamma =0.5$) power-law exponents of the potential term, respectively. 
The solid line is the result of a linear interpolation between the predictions, weighted according to the experimental errors of 
the included four bins in $p_{t}/A$, and leading to the 
indicated $\gamma =0.75 \pm 0.10$
(from Ref.~\protect\cite{russotto16}; 
Copyright (2016) by the American Physical Society).
}
\label{fig:diffdata}       
\end{figure}

The predictions obtained with these assumptions for the measured flow ratio are shown in Fig.~\ref{fig:diffdata} together with the experimental
result. The histogram represents the linear interpolation between the predictions giving a best fit of the flow ratios presented here as a function of the
transverse momentum $p_t$. The corresponding power-law coefficient is $\gamma = 0.75 \pm 0.10$. In comparison with FOPI-LAND, the statistical 
accuracy $\Delta \gamma = 0.10$ represents a strong improvement by more than a factor of two.
A systematic uncertainty arose from occasional malfunctions of the electronic circuits for the time measurement with LAND. 
It prohibits extending the analysis into the weakly polulated region with $p_t > 0.7$~MeV/$c$. It is of considerably less importance for the flow ratio 
constructed after integrating the measured yields over the full acceptance of LAND in the $p_t$-vs-rapidity plane. Only a minor uncertainty remains that
is related to the lower energy threshold of neutron detection. The necessary corrections and the methods used for estimating the resulting errors are
described in detail in Ref.~\cite{russotto16}. They include, e.g., a small correction caused by the possible misinterpretation of neutrons as charged 
particles and vice versa, caused by reactions of neutrons in the veto wall and by the narrow gaps between the veto-wall paddles.  

With all corrections and errors included, the acceptance-integrated elliptic-flow ratio leads to a power-law coefficient $\gamma = 0.72 \pm 0.19$. 
This is the result displayed in Fig.~\ref{fig:final} as a function of the reduced density $\rho/\rho_0$. The new result confirms the former and has a 
considerably smaller uncertainty. It is also worth noting that the present parametrization is compatible with the low-density behavior of the symmetry 
energy from Refs.~\cite{brown13,zhang13,tsang09,dani14} that are included in the figure.  
The corresponding slope parameter describing the variation of the symmetry energy with density at saturation
is $L = 72 \pm 13$~MeV. The sharp value $E_{\rm sym} (\rho_0) = 34$~MeV is a consequence of the chosen parametrization (Eq.~\ref{eq:pot_term}). 
Using values lower than the default $E_{\rm sym}^{\rm pot} (\rho_0) = 22$~MeV, as occasionally done in other UrQMD 
studies~\cite{wang14a}, will lower the result for $L$. Performing the present UrQMD analysis with $E_{\rm sym}^{\rm pot} (\rho_0) = 19$~MeV, 
corresponding to $E_{\rm sym} (\rho_0) = 31$~MeV, yields a power-law coefficient $\gamma = 0.68 \pm 0.19$ and a slope 
parameter $L = 63 \pm 11$~MeV. The observed changes remain both within the error margins of these quantities. However,
the precise results of Brown~\cite{brown13} and Zhang and Chen~\cite{zhang13} are not equally 
met with this alternative parametrization of the symmetry energy.

\section{Sensitivity to density}
\label{sec:density}

The sensitivity to density was explored in a study using the T\"{u}bingen version~\cite{cozma13} of the QMD model (T\"{u}QMD). The underlying idea
consisted in performing transport calculations for the present reaction with two parametrizations of the symmetry energy that were chosen
to be different for a selected range of density and identical elsewhere. The magnitude of the obtained difference between the two 
predictions for the elliptic flow ratio is considered a measure of the sensitivity to the selected density region.

In the momentum-dependent one-body potential introduced by Das {\it et al.}~\cite{Das:2002fr}, the stiffness of the symmetry energy
is controlled with a parameter $x$ and choices ranging from soft ($x=+1$) up to rather stiff ($x=-2$) density dependences 
are commonly selected in model studies.
For the present case, a mildly stiff ($x=-1$) and a soft ($x=+1$) parametrization were chosen for the density range with different 
symmetry energy while the nearly linear case with $x=0$ was chosen for the common part. 

To quantify the results, a quantity DEFR (Difference of Elliptic-Flow Ratio) was defined as
\begin{eqnarray}
{\rm DEFR}^{(n,Y)}(\rho)=\frac{v_2^n}{v_2^{Y}}(x=-1,\rho)-\frac{v_2^n}{v_2^{Y}}(x=1,\rho).
\label{defrdef}
\end{eqnarray}
It represents the differences of the elliptic flow ratios calculated with the T\"{u}QMD transport model for two different 
density dependencies of the symmetry energy. Here $Y$ indicates a particle or a group 
of particle species and $x$ the stiffness that is used in the calculations at densities smaller than the argument $\rho$. 
The difference of the $x = \pm 1$ potentials is thus effectively only tested
at densities up to the particular $\rho$, the argument of DEFR. This choice leads 
to DEFR$^{(n,Y)} (0) = 0$ and to the full stiff-soft splitting for large values of the argument $\rho$. The slope of DEFR
at intermediate densities is a measure of the impact on elliptic flow observables of that particular 
region of density.

\begin{figure}[t]           
\centering
\includegraphics[width=7.5cm]{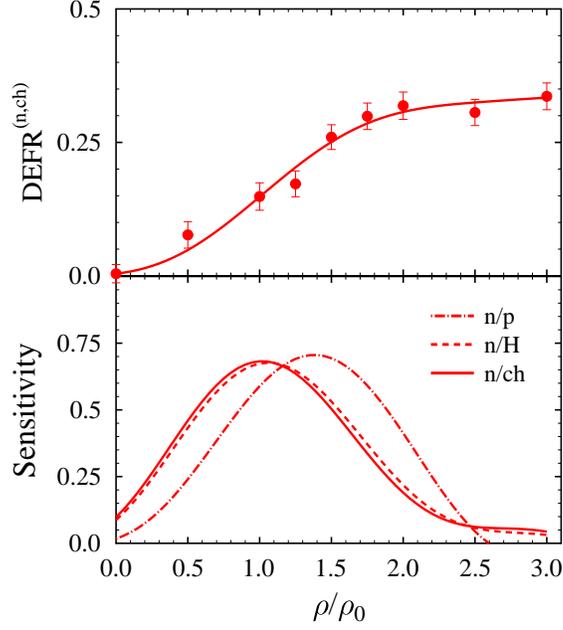}
\caption{Density dependence of the difference of the elliptic flow ratio (DEFR) of neutrons
over charged particles, defined by Eq.~\protect\ref{defrdef}, for $^{197}$Au+$^{197}$Au collisions 
at 400 MeV/nucleon obtained with the T\"{u}QMD transport model~\protect\cite{cozma13} and 
the FOPI-LAND acceptance filter (top) and the corresponding sensitivity density (bottom panel, solid line) together with
sensitivity densities obtained from elliptic-flow ratios of neutrons over all hydrogen isotopes (dashed) and 
neutrons over protons (dash-dotted; 
from Ref.~\protect\cite{russotto16}; 
Copyright (2016) by the American Physical Society).
}
\label{fig:dens}       
\end{figure}

In the upper panel of Fig.~\ref{fig:dens}, the density dependence of DEFR$^{(n,Y)}$ for the choice $Y$=all charged 
particles is presented. It is seen that DEFR increases monotonically up to density values in the neighborhood 
of 2.5\,$\rho_0$, close to the maximum density probed by nucleons in heavy-ion collisions at 400 MeV/nucleon 
incident energy. The distribution of the sensitivity as a function of density is obtained by forming the 
derivative of DEFR with respect to density.
It is presented in the lower panel of Fig.~\ref{fig:dens} for three choices of $Y$, only protons (n/p), sum of all 
hydrogen isotopes (n/H), and all charged particles (n/ch).
 
The figure shows that the sensitivity achieved with the elliptic-flow ratio of neutrons over charged particles, 
the case studied here, reaches its maximum close to saturation density and extends beyond twice that value. It is compatible
with the conclusions reached by Le~F\`{e}vre {\it et al.} in their study of the symmetric matter EoS, based on FOPI 
elliptic-flow data and calculations with the Isospin Quantum Molecular Dynamics transport model~\cite{lefevre16}.
For $^{197}$Au+$^{197}$Au collisions at 400 MeV/nucleon, the broad maximum of the force-weighted density defined by the authors
is spread over the density range $0.8 < \rho /\rho_0 < 1.6$.

The sensitivity of the neutron-vs-proton flow ratio has its maximum in the 1.4 to 1.5 $\rho_0$ region, i.e. at 
significantly higher densities than with light complex particles being included (Fig.~\ref{fig:dens}).
This observation contains an important potential for future experiments. With efficient isotope separation, 
flow measurements may give access to the curvature of the symmetry energy at saturation, in addition to the slope. 
It is, therefore, of extreme importance for future experiments to be able to extract
a clean separate proton signal. Theoretical models to be used in the analysis should permit the
independent adjustment of the slope and curvature parameters of the symmetry-energy term. With improvements in these directions and
higher beam energies, extracted constraints for the EoS of asymmetric matter can be expected to reach into the 2$\rho_0$ regime.

\section{Conclusion and Outlook}

The sensitivity study shows that suprasaturation densities are effectively probed with the elliptic flow ratio of neutrons with respect to charged particles. 
Because the interpretation of the FOPI pion ratios~\cite{reisdorf07} is not yet conclusive (see, e.g., Refs.~\cite{hong14,xiao14,cozma16}), 
this observable is presently unique as a terrestrial source of information for the EoS of asymmetric matter at high densities.

The value $\gamma = 0.72 \pm 0.19$ obtained for the power-law coefficient of the potential part in the UrQMD parametrization of the symmetry energy 
and the slope parameter $L = 72 \pm 13$~MeV are equivalent to a symmetry pressure 
$p_0 = \rho_0 L/3 = 3.8 \pm 0.7$~MeVfm$^{-3}$. The latter may be used to estimate the pressure in neutron-star matter at saturation density. For an assumed
asymmetry $\delta = (\rho_n - \rho_p)/\rho = 0.9$ in that part of the star, it amounts to 3.4~MeVfm$^{-3}$ (Ref.~\cite{russotto16}), a value that compares well with the pressure obtained by 
Steiner {\it et al.}~\cite{steiner13} from neutron-star observations.
  
Regarding the modeling of nuclear reactions,
it will be important to improve the description of the nuclear interaction in transport models,
to reduce the parameter ranges also in the isoscalar sector, to improve the algorithms used for clusterization,
as well as going beyond the mean-field picture, including short-range correlations. 
The latter have recently been investigated in nuclei~\cite{subedi08,hen14sc} and their consequences 
for transport descriptions of intermediate-energy heavy-ion reactions need to be investigated.
Moreover, it will be quite important to compare the experimental data with the predictions of several transport models, 
of both Boltzmann-Vlasov and molecular-dynamics type~\cite{guoyongwang13}, in order to pursue the work towards a 
model-independent constraint of the high-density symmetry energy initiated in Refs.~\cite{russotto_epja14,cozma13}.

The presented experimental results and the theoretical study of the density range that has been probed, provide a strong 
encouragement for continuing flow measurements of the present kind with improved detection systems. Preliminary model 
studies indicate that the sensitivity to the stiffness of the symmetry energy is still significant at incident 
energies as high as 800 MeV or even 1 GeV/nucleon. 
The density study suggests that the curvature parameter $K_{\rm sym}$ can be addressed experimentally
if higher precision and elemental and isotopic resolution for charged particles can be achieved.  
Future experiments will, therefore, benefit from the improved capabilities of the NeuLAND detector~\cite{NeuLAND} 
presently constructed as part of the $R^{3}B$ setup for experiments at FAIR. Continuing experimental 
and theoretical activities can thus be expected to drive the range of densities that can be probed in the laboratory up to twice saturation.  

The results presented in this talk have been obtained within the ASY-EOS Collaboration. See Ref.~\cite{russotto16} for the complete list of authors.

\end{document}